\documentclass[12pt, epsfig]{article}                            
\usepackage{epsfig}       
\usepackage{amssymb}     
\usepackage{amsfonts}     
\newskip\humongous \humongous=0pt plus 1000pt minus 1000pt

\newif\ifdtup

\catcode`\@=11
 
\@addtoreset{equation}{section}
\def\theequation{\thesection.\arabic{equation}}
 
\def\@normalsize{\@setsize\normalsize{15pt}\xiipt\@xiipt
\abovedisplayskip 14pt plus3pt minus3pt%
\belowdisplayskip \abovedisplayskip
\abovedisplayshortskip \z@ plus3pt%
\belowdisplayshortskip 7pt plus3.5pt minus0pt}
 
\def\small{\@setsize\small{13.6pt}\xipt\@xipt
\abovedisplayskip 13pt plus3pt minus3pt%
\belowdisplayskip \abovedisplayskip
\abovedisplayshortskip \z@ plus3pt%
\belowdisplayshortskip 7pt plus3.5pt minus0pt
\def\@listi{\parsep 4.5pt plus 2pt minus 1pt
     \itemsep \parsep
     \topsep 9pt plus 3pt minus 3pt}}
 
\relax

\catcode`@=12
 
\catcode`\@=11
 
\def\section{\@startsection{section}{1}{\z@}{3.5ex plus 1ex minus
   .2ex}{2.3ex plus .2ex}{\large\bf}}
 
\def\thesection{\arabic{section}}    
\def\thesubsection{\arabic{section}.\arabic{subsection}}

\def\appendix{\setcounter{section}{0}
 \def\thesection{Appendix \Alph{section}}
 \def\thesubsection{\Alph{section}.\arabic{subsection}}
 \def\theequation{\Alph{section}.\arabic{equation}}}
\def\YGrule{0.4}   % line thickness in unit of pt
\def\YGbox{6.5}    % box size in unit of pt
\def\SymBoxes#1#2#3#4{\newdimen\un@t \un@t#3%
\raisebox{#1}{\rule{#2\un@t}{#4}\hskip-#2\un@t% lower horizontal
\@tempdimb\un@t \advance\@tempdimb by-#4\@tempcntb#2\relax%
\@whilenum{\@tempcntb>0}\do{%                         % #2 vertical lines
\rule{#4}{\un@t}\hskip\@tempdimb \advance\@tempcntb by\m@ne}%
\hskip-#2\un@t \rule[\un@t]{#2\un@t}{#4}%
\rule[\un@t]{#4}{#4}\hskip-#4%             % upper horizontal line
\rule{#4}{\un@t}}\hskip-#4}                % rightest vertical line
\def\Young{\@ifnextchar[{\@Young}{\@Young[0]}}
\def\@Young[#1]#2{\newdimen\YG@unit \YG@unit\YGbox pt%
\newdimen\h@ight \h@ight#1\YG@unit \@tempcnta-1\relax
\@tfor\c@ount:=#2\do{\advance\@tempcnta by\@ne}% count the number of rows
\@tempdima\@tempcnta\YG@unit%
\advance\h@ight by\@tempdima\relax     % compute the height of the top row
\@tfor\c@ount:=#2\do{\SymBoxes{\h@ight}{\c@ount}{\YG@unit}{\YGrule pt}%
\@tempdima-\c@ount\YG@unit \hskip\@tempdima%
\advance \h@ight by -\YG@unit}         % Draw the Tableaux
\@tempdima\YG@unit \multiply\@tempdima by\@car#2\@nil %
\hskip\@tempdima}                      % hskip by the length of the top row
\def\YoungTab{\@ifnextchar[{\@YoungIdx}{\@YoungIdx[0]}}
\def\@YoungIdx[#1]{\@ifnextchar[{\@iYoungIdx[#1]}{\@iYoungIdx[#1][\@empty]}}
\def\@iYoungIdx[#1][#2]#3{%
\newdimen\YG@unit \YG@unit\YGbox pt\newdimen\YG@rule \YG@rule \YGrule pt
\newcount\c@ount \c@ount\z@ \newdimen\skip@wd \unitlength\@ne pt
\newdimen\h@ight \h@ight#1\YG@unit \@tempcnta\m@ne\relax
\@tfor\d@um:=#3\do{\advance\@tempcnta by\@ne}% count the number of rows
\@tempdima\@tempcnta\YG@unit%
\advance\h@ight by\@tempdima\relax%  % compute the height of the top row
\@tfor\@idxlist:=#3\do{%             % routine to draw the indexed Tableaux
\@tempcnta\z@\hskip.5\YG@rule\relax 
\@for\@idx:=\@idxlist\do{%           % place the indices of the row first
\raisebox{\h@ight}{\makebox(\YGbox,\YGbox){#2$\@idx$}}
\advance\@tempcnta by\@ne}\hskip-.5\YG@rule% 
\@tempdima-\@tempcnta\YG@unit \hskip\@tempdima%
\ifnum\c@ount=\z@ \skip@wd-\@tempdima\fi \relax% record the top row width
\SymBoxes{\h@ight}{\@tempcnta}{\YG@unit}{\YG@rule}%
\hskip\@tempdima \advance\h@ight by -\YG@unit
\advance\c@ount by\@ne}%             % end of the routine
\hskip\skip@wd}                      % hskip by the length of the top row
\begin{document}
%\begin{letter}{~}

%%%%%%Define some new commands and  macros
\newcommand{\beq}{\begin{equation}}
\newcommand{\eeq}{\end{equation}}
\newcommand{\bea}{\begin{eqnarray}}
\newcommand{\eea}{\end{eqnarray}}
\newcommand{\beas}{\begin{eqnarray*}}
\newcommand{\eeas}{\end{eqnarray*}}
\newcommand{\defi}{\stackrel{\rm def}{=}}
\newcommand{\non}{\nonumber}
\newcommand{\bquo}{\begin{quote}}
\newcommand{\enqu}{\end{quote}}
%%%%%%%%%%%%%%%%%%%%%%%%%%%%%%%%%% definitions
\def\de{\partial}
\def\Tr{ \hbox{\rm Tr}}
\def\const{\hbox {\rm const.}}
\def\o{\over}
\def\im{\hbox{\rm Im}}
\def\re{\hbox{\rm Re}}
\def\bra{\langle}\def\ket{\rangle}
\def\Arg{\hbox {\rm Arg}}
\def\Re{\hbox {\rm Re}}
\def\Im{\hbox {\rm Im}}
\def\diag{\hbox{\rm diag}}
\def\longvert{{\rule[-2mm]{0.1mm}{7mm}}\,}
\begin{titlepage}
{\hfill     IFUP-TH/2003-50} 
\bigskip
\bigskip

\begin{center}
{\large  {\bf  
Nonabelian Monopoles and the Vortices that Confine Them
 } } 
\end{center}

\bigskip
\begin{center}
{\large  Roberto AUZZI $^{(1,3)}$ , Stefano BOLOGNESI $^{(1,3)}$, \\
 Jarah EVSLIN $^{(3,2)}$ and   Kenichi KONISHI $^{(2,3)}$
 \vskip 0.10cm
 }
\end{center}

\begin{center}
{\it   \footnotesize
Scuola Normale Superiore - Pisa $^{(1)}$,
 Piazza dei Cavalieri 7, Pisa, Italy \\
Dipartimento di Fisica ``E. Fermi" -- Universit\`a di Pisa $^{(2)}$, \\
Istituto Nazionale di Fisica Nucleare -- Sezione di Pisa $^{(3)}$, \\
     Via Buonarroti, 2, Ed. C, 56127 Pisa,  Italy $^{(2,3)}$ 
   }

\end {center}

\noindent  
{\bf Abstract:}

Nonabelian magnetic monopoles  of Goddard-Nuyts-Olive-Weinberg type  have recently been  shown to appear as the dominant
infrared  degrees of freedom in a class of  softly broken ${\cal N}=2$  supersymmetric gauge theories in which the gauge group $G$  is
broken  to various  nonabelian subgroups  $H $ by an adjoint Higgs VEV.
When the low-energy gauge group $H$ is further broken  completely by e.g.  squark VEVs,   
the monopoles  representing  $\pi_2(G/H)$  are confined by  the nonabelian vortices  arising from the breaking of 
$H$,  discussed recently   
(hep-th/0307278).  By considering the system with  $G=SU(N+1)$,  $ H = {SU(N) \times U(1) \o {\mathbb Z}_N}$,   as an example,   we show   that the total  
magnetic flux of  the minimal  monopole  agrees   precisely with the total magnetic flux flowing along the single minimal vortex.
 The possibility for 
such an analysis reflects the presence of free   parameters in the theory - the bare quark mass $m$ and the adjoint mass $\mu$ - such  
that for $m \gg \mu$   the topologically nontrivial solutions of vortices and of unconfined monopoles exist at distinct energy scales.

\vfill  
 
\begin{flushright}
  December    2003
\end{flushright}
\end{titlepage}

\bigskip

\hfill{}
\bigskip

\section{Introduction }

Nonabelian monopoles in spontaneously broken gauge theories have remained  somewhat   obscure 
objects for a long time in spite of many investigations \cite{Lb}-\cite{CJH}.          Apart from the often discussed applications in conformally invariant
${\cal N}=4$ theories,  few field theory models  were   known where such objects play an important dynamical role.   Although 
many ${\cal N}=1$  gauge theories, such as SQCD with appropriate numbers of flavors,    are believed to possess Seiberg  duals
\cite{Sei},    the origin of  the   ``dual quarks" appearing in these models   remains mysterious.
 
 A series of papers on softly broken ${\cal N}=2$  gauge theories with gauge groups $SU(N)$, $USp(2N)$ and
$SO(N)$ and various numbers of flavors of fundamental matter have,   however, changed the situation \cite{APS,CKM,CKKM}.    In particular, it was
pointed out
\cite{BK} that the ``dual quarks" appearing as the low-energy degrees of freedom of the $G= SU(N),$ $USp(2N)$ or    
$SO(N)$  theory,  which carry the    nonabelian 
$SU(r) \subset G$  charges,   can  be identified with  the ``semiclassical" nonabelian monopoles studied earlier by  Goddard, Nuyts, Olive \cite{GNO}
and  by E. Weinberg \cite{EW}.   Also, all  of the confining vacua in strongly coupled  $USp(2N)$ and $SO(N)$ theories with  flavors
and with zero bare quark masses,   involve these objects in a deformed SCFT.

Very  recently, with A. Yung, we have proven the existence of   nonabelian {\it vortices }     in the
same class of models \cite{ABEKY}.   The analysis was done semiclassicaly,   in the  region of large   bare quark masses (and so large
adjoint scalar VEVS),  but  the presence of an appropriate number of fermions makes the results quantum mechanically correct.  In particular, a continuous
family of degenerate vortex solution  have been  constructed, showing the truely nonabelian nature of these vortices \footnote{Deceptively
similar, though different,   vortex configurations  have been studied independently by  Hanany and Tong \cite{HT,Tong}.}.

In this paper,  we  discuss some aspects {\it relating }       nonabelian vortices and monopoles appearing in the softly broken ${\cal N}=2$  $G=SU(N)$  theories
with
$N_f$  flavors.   The gauge group is broken at two very different mass scales,   $v_1 \gg v_2$, 
\beq    G   \,\,\,{\stackrel {v_1}{\longrightarrow}}     \,\,\, H     \,\,\,{\stackrel {v_2}{\longrightarrow}}     \,\,0. 
\eeq
For concreteness we shall study  the case of the symmetry breaking   $G=SU(N+1), $  $ H = {SU(N) \times U(1) \o {\mathbb Z}_N}$:  the symmetry breaking at
the   higher mass scale is due to the adjoint scalar VEV which is proportional to the bare quark masses  $v_1 \sim m$   (see Eq.(\ref{adjvev}) below);    the
squark VEVS  break
$H$ at much smaller mass scale, $v_2 \sim \sqrt{\mu  m}$,  where $\mu$  is the small adjoint scalar mass, breaking the supersymmetry to ${\cal N}=1.$
 The model is basically   the bosonic sector of the 
${\cal N}=2$  supersymmetric gauge  theories \cite{SW1}-\cite{curves}.   
The full supersymmetric dynamics of the theory involving fermions is however  needed  to show the
quantum mechanical  stability of  what is found  here semiclassically \cite{BK,ABEKY}.

Strictly speaking, neither the monopoles nor  vortices exist in this  theory as static, topologically stable configurations, 
as  $\pi_2(G)= \pi_1(G)=0.$  Nonetheless, the existence of the  different scales  in the theory allows us to study these configurations 
as approximate,  topologically stable static configurations of effective theories defined  at  different scales.  
At high energies where the effects of the smaller condensates are negligible,  the theory possesses the nonabelian monopoles representing 
nontrivial elements of $\pi_2(G/H)$  and transforming as multiplets of the dual gauge group ${\tilde H}$.       At lower energies the light fields are described by
an effective
$H$ theory
 in the Higgs phase.  This theory possesses  vortices of  $\pi_1(H)$ which are stable  in so far   as the  pair production
 of the massive monopoles is  suppressed.

The equivalence of the homotopy groups   $\pi_2(G/H)\sim  \pi_1(H)  $   implies  that  the monopoles are confined.  
 Although the configuration  of an isolated  monopole has an infinite energy in the Higgs phase ($v_2 \ne 0$),   the flux of the monopole
can be  whisked   away by a single  vortex,  so that a  monopole-vortex-antimonopole configuration has a finite energy. 
{\it In making this discussion more quantitative, we show  that the flux through a small sphere around a  monopole  exactly matches
the flux along the vortex through  a plane perpendicular  to it, far from the monopole.   }   

Although this result is  in a sense to be   expected,  it is actually a quite nontrivial matter  to show it, as in the present model the monopoles are
``made of"  gauge bosons and adjoint scalars while  the vortices are  nontrivial configurations involving gauge fields and squarks  fields only;   the two 
types of  configurations appear as solutions of two different  effective theories valid at different  energy scales.

 More importantly,  this discussion  shows   that the
 monopoles  indeed  form   a  nonabelian (dual) gauge multiplet,    since    continuous transformations of the vortex solutions
have recently been explicitly constructed \cite{ABEKY},   proving  their nonabelian nature.

\section{Nonabelian Monopoles and Vortices in $SU(N)$  Gauge Theories \label{NAMonopoles}}

\subsection {High-energy theory:  BPS monopoles}

%General properties of the nonabelian magnetic monopoles 
%and explicit formulas are summarized   in Appendix A. 
 We start our discussion by considering the monopoles and vortices arising in a  
system with symmetry breaking $SU(N+1) \to SU(N) \times U(1)$.  The field theory considered here  is essentially  the bosonic sector of
${\cal N}=2$ supersymmetric $SU(N+1)$     gauge theory \cite{curves}. 
     The discussion of this section is semi-classical,
although when 
 embedded in the ${\cal N}=2 $ theory and with appropriate number of flavors (in this case, $2N \le  N_f   \le     2N+2$),  
 the whole discussion is valid quantum
mechanically.

For concreteness, in this subsection we discuss  the   ${\cal N}=2,$  $SU(3)$ gauge theory with $n_{f}=4,5$   flavors of hypermultiplets (``quarks'').
The generalization to systems with more general pattern of symmetry breaking $SU(N+1) \to SU(r) \times U(1)^{N-r+1}$   is straightforward.   
The results for the monopole-vortex flux   matching in the next section will be given for $SU(N+1) \to SU(N) \times U(1)$  cases as well.

The Lagrangian of this theory has the structure
\beq
{\cal L}=     {1\over 8 \pi} \im \, S_{cl} \left[\int d^4 \theta \,
\Phi^{\dagger} e^V \Phi +\int d^2 \theta\,{1\o 2} W W\right]
+ {\cal L}^{(quarks)}  +  \int \, d^2 \theta \,\mu  \,\Tr  \Phi^2;  
\label{lagrangian}
\eeq
\beq {\cal L}^{(quarks)}= \sum_i \, [ \int d^4 \theta \, \{ Q_i^{\dagger} e^V
Q_i + {\tilde Q_i}  e^{-V} {\tilde Q}_i^{\dagger} \} +  \int d^2 \theta
\, \{ \sqrt{2} {\tilde Q}_i \Phi Q^i    +      m   {\tilde Q}_i Q^i   \}
\label{lagquark}
\eeq
where $m$ is the bare mass of the quarks and we have defined the complex coupling constant
\beq
S_{cl} \equiv  {\theta_0 \o \pi} + {8 \pi i \o g_0^2}.  
\label{struc}
\eeq   
The parameter $\mu$ is the mass of the adjoint chiral multiplet, which breaks the supersymmetry to ${\cal N}=1$.

In order to discuss unconfined monopoles, however,    we must set $\mu=0$ (see Subsec. \ref{incomp} below) and so preserve the full ${\cal N}=2$
supersymmetry. 
     After eliminating the auxiliary fields  the bosonic Lagrangian becomes 
\beq  {\cal L}=  { 1\o 4 g^2}  F_{\mu \nu}^2  +  { 1\o g^2}  |{\cal D}_{\mu} \phi|^2 +  
 \left|{\cal D}_{\mu}
Q\right|^2 + \left|{\cal D}_{\mu} \bar{\tilde{Q}}\right|^2 +  {\cal L}_1+  {\cal L}_2, 
\label{Lag}\eeq
where
\bea    {\cal L}_1 &=&  -   { 1\o 8 }  \, \sum_A  [ { 1\o g^2 }  (-  i)  f_{ABC} \, \phi^{\dagger}_B  \phi_C +  Q^{\dagger} t^A  Q -  {\tilde Q} t^A   {\tilde
Q}^{\dagger} ]^2    \non \\
&=&     -   { 1\o 8 }  \, \sum_A  \left( t^A_{ij} \,  [ { 1\o g^2 }  (-2)  \, [\phi^{\dagger},   \phi]_{ji}  +  Q^{\dagger}_j   Q_i -  {\tilde Q}_j   {\tilde
Q}^{\dagger}_i ]^2 \right)^2; 
\eea
\bea    {\cal L}_2 &=&   - g^2 |   \mu \, \phi^A +
\sqrt 2   \, {\tilde Q} \, t^A  Q |^2   -   {\tilde Q } \,   [ m    + \sqrt2  \phi   ] \,  [ m    + \sqrt2  \phi   ]^{\dagger}  \, {\tilde Q}^{\dagger} 
\non \\  &-& Q^{\dagger} \, [ m    + \sqrt2  \phi   ]^{\dagger}    \, [ m    + \sqrt2  \phi   ]  \, Q.  
\eea
In the construction of the monopole solutions  we shall consider only the VEVs and fluctuations around them which satisfy 
\beq   [\phi^{\dagger},   \phi]=0, \qquad    Q_i =  {\tilde Q}^{\dagger}_i,
\eeq
and hence   ${\cal L}_1$,  ${\cal L}_2$  can be set identically to zero. 

This theory has a number of vacua parametrized by the integer $r$, which is the rank of the unbroken nonabelian gauge symmetry plus one \cite{APS,CKM}.  
For concreteness we first consider the $r=2$ vacuum, in which
 the adjoint scalar has a nonvanishing VEV  ($\Phi=  t^a  \phi^a$)
\beq  diag. \bra \Phi \ket =    v_1\,  (1, 1,   -  2 )  ;  \quad  \bra  \phi^b \ket =0, \quad b=1,2,3,  \quad  
\bra  \phi^8 \ket =-
2\sqrt {3}\,  v_1, 
\eeq
while the squark VEVs are set to zero, $Q_i =  {\tilde Q}^{\dagger}_i=0$.   We will consider the semiclassical regime in which the bare quark mass is much larger than the QCD scale, in which case $v_1=m/\sqrt{2}$ and so
\beq  diag. \bra \Phi \ket =    { 1\o \sqrt 2}    (m, m,   -  2 \, m);     \quad  
\bra  \phi^8 \ket =-
\sqrt {6}m. \label{adjvev}\eeq
This VEV breaks the gauge symmetry as 
\beq   SU(3)  \to   {SU(2)  \times U(1)  \o {\mathbb Z}_2 },    
\label{Symbr}\eeq
where the $ {\mathbb Z}_2 $  factor arises because $SU(2) $ and $  U(1)$ share  the common element $- {\bf 1}$. 

The nontrivial homotopy groups  
\beq \pi_2({SU(3)\o  SU(2) \times U(1) / {\mathbb Z}_2})  = \pi_1(SU(2) \times U(1) / {\mathbb Z}_2) =  {\mathbb Z} \label{homotopy}  \eeq 
imply that nontrivial  monopole solutions exist.  The energy of such configurations may be read from the Hamiltonian  
    \beq  H = \int d^3x  \, \Big[   { 1\o 4 g^2}  (F_{i j}^A)^2  +  { 1\o g^2}  |{\cal D}_{i} \phi^A|^2  \Big ] = \int d^3x  \, \Big[   { 1\o 4 g^2}  (F_{i
j}^A)^2  +  { 1\o 2   g^2}  |{\cal D}_{i} \phi^A|^2  \Big ] 
\eeq
where   in the second formula we have kept only the real part of $\phi^A$.  Note that we have restricted our interest to static configurations with no electric flux.  For  real  $\phi^A$,   $f_{ABC} \, \phi^{\dagger}_B  \phi_C=0$  so neither 
${\cal L}_1$ nor ${\cal L}_2$  contribute.  Rewriting the Hamiltonian as      
\beq   H=     \int d^3x  \, \Big[   { 1\o 4 g^2} |  F_{i j}^A  \pm \epsilon_{ijk}     ({\cal D}_{k}
\phi)^A  |^2  \pm { 1\o 2}  \de_k  (\epsilon_{ijk}   F_{i j}^A 
\phi^A)\Big ]
\eeq
it becomes clearer that BPS monopole configurations must satisfy the nonabelian Bogomolny equations
\beq  B_k^A= -   ({\cal D}_{k} \phi)^A; \qquad  B_k^A = {1 \o 2}\,  \epsilon_{ijk}  F_{i j}^A \label{nbe} . 
\eeq
The BPS bound on the monopole mass is then (see Eq.(\ref{su3sol}), Eq.(\ref{calcu}), Eq.(\ref{so3flux})  below)
\beq  H=  \int dS \cdot  (\phi^A  {\bf B}^A ) = { 2 \pi \o g} \,3  \, v_1 \, m, \qquad m =1,2, \ldots.
\eeq

\subsection{Low-energy theory:  Vortices}   

Vortices appear in the low-energy theory when the symmetry group $ {SU(2)  \times U(1)  \o {\mathbb Z}_2 }   $  is further   spontaneously broken by squark
VEVs \cite{ABEKY}.    Upon turning on an adjoint mass perturbation ($\mu \ne 0$),  the squark VEVs  take a color-flavor diagonal form  ($\xi \equiv \mu \, m$): 
\beq
\label{qvev}
<q^{kA}>=<\bar{\tilde{q}}^{kA}>=\sqrt{\frac{\xi}{2}}\left(
\begin{array}{cc}   1 & 0  \\   0 & 1  \\
  \end{array}\right)  =   v_2 \, \left(
\begin{array}{cc}   1 & 0  \\   0 & 1  \\
  \end{array}\right),     \eeq 
where only the first two color and flavor components are  explicitly shown   (all other components being identically zero in the vortex solution).  
  The light fields  enter the  $SU(2) \times  U(1)$ Lagrangian at scales  between  $v_1 $ and $v_2$   (we set ${\cal L}_1=0$) as 
\bea  {\cal L} &=&  { 1\o 4 g_2^2}  (F_{\mu \nu}^a)^2  + { 1\o 4 g_1^2}  (F_{\mu \nu}^0)^2  +  { 1\o g_2^2}  |{\cal D}_{\mu} \phi^a |^2 + 
 { 1\o g_1^2}  |{\cal D}_{\mu} \phi^0 |^2 +   
 \left|{\cal D}_{\mu}
Q \right|^2 + \left|{\cal D}_{\mu} \bar{\tilde{Q}}\right|^2  \non \\
&-&    g_2^2 |  \,  \mu \, \phi^8 +
\sqrt 2   \, {\tilde Q} \, t^8  Q \, |^2 - g_1^2 |  
\sqrt 2   \, {\tilde Q} \, t^a Q \,  |^2   -   {\tilde Q } \,   [ m    + \sqrt2  \phi   ] \,  [ m    + \sqrt2  \phi   ]^{\dagger}  \, {\tilde Q}^{\dagger} 
\non \\  &-& Q^{\dagger} \, [ m    + \sqrt2  \phi   ]^{\dagger}    \, [ m    + \sqrt2  \phi   ]  \, Q,  
\eea
where $a=1,2,3$ labels the $SU(2)$ generators,  $t^a=  S^a $;   the index $0$ refers to $t^8= { 1 \o 2 \sqrt{3}} \, \diag (1,1, -2).$    We have taken into
account the different renormalization effects in the $SU(2)$ sector and $U(1)$ sector and  distinguished the coupling constants $g_2$ (of $SU(2)$ interactions) 
and 
$g_1$  (of $U(1)$).

Note that  the model discussed by Hanany and Tong \cite{HT,Tong}  is different, as  the FI term in the $U(1)$ part is put in by hand,
while in our model  the corresponding term is an F term, arising naturally from the $SU(3) \to SU(2) \times U(1)$ breaking.   Also, our monopoles and  vortices
have quantum mechanical meaning as the  $SU(2) \times U(1)$   is  infrared free in the scales between 
$v_1$ and $v_2$.  While they have found two vortices ending on each monopole in their model, we will see that the monopoles of our model are each confined by a single vortex.

The static field   energy of an arbitrary configuration without electric flux is
\bea  H  &=&  \int d^3x  \, \Big[ \,  { 1\o 4 g_2^2}  (F_{ij  }^a)^2  + { 1\o 4 g_1^2}  (F_{ij }^0)^2   +  { 1\o g_2^2}  |{\cal D}_i  \phi^a|^2 + 
{ 1\o g_1^2}  |{\cal D}_i  \phi^0|^2 + 
 \left|{\cal D}_i Q \right|^2 + \left|{\cal D}_i \bar{\tilde{Q}}\right|^2 +   \non \\
&+&    g_2^2 |  \,  \mu \, \phi^8 +
\sqrt 2   \, {\tilde Q} \, t^8  Q \, |^2  +  g_1^2 |  
\sqrt 2   \, {\tilde Q} \, t^a Q \,  |^2   +     {\tilde Q } \,   [ m    + \sqrt2  \phi   ] \,  [ m    + \sqrt2  \phi   ]^{\dagger}  \, {\tilde Q}^{\dagger} 
\non \\  &+  & Q^{\dagger} \, [ m    + \sqrt2  \phi   ]^{\dagger}    \, [ m    + \sqrt2  \phi   ]  \, Q  \Big] . 
\eea
Now let us retrict our attention to those configurations in which  the adjoint scalar is fixed  to its VEV, 
\beq   \phi =  v_1 \,   t^8, 
\eeq
which is constant and commutes with  $t^a$ and $t^8$ and also satisfies   $ {\cal D}_i  \phi^a \to 0.$
By also keeping   $Q= {\tilde Q}^{\dagger} $, rescaling $Q =  { 1\o \sqrt 2}  q $,  and 
keeping the  first two color and flavor components of  these to be nonvanishing,  
 one obtains the Hamiltonian
\bea  H  %&=&  \int d^3x  \, \Big[  \,  { 1\o 4 g_2^2}  (F_{ij  }^a)^2  + { 1\o 4 g_1^2}  (F_{ij }^0)^2   + 
 %\left|{\cal D}_i q \right|^2   \non \\
%&+&    g_1^2 \, |  - \sqrt 6 \, m  \, \mu \, +
%{1 \o 2 \sqrt 6}    \, q^{\dagger} \, q  \, |^2  +  g_2^2 \, |  
%{1 \o 2 \sqrt 2}    \, q^{\dagger}  \, S^a q \,  |^2   \Big]  \non \\
&=&    \int d^3x  \,  \Big [   
  |{ 1\o 2 g_1} F_{ij  }^0 \pm \epsilon_{ij}   { g_1 }( - \sqrt 3 \, m  \, \mu \, +
{1 \o 4\sqrt 3}    \, q^{\dagger} \, q  \, )   |^2  + \non \\
&+& |{ 1\o 2 g_2} F_{ij  }^a  \pm \epsilon_{ij}   { g_2 \o 4} \, q^{\dagger}  \, S^a q \, |^2 + 
\frac{1}{2} \left|{\cal D}_i \,q^A \pm   i   \epsilon_{ij}
{\cal D}_j\, q^A\right|^2
\pm
2  \sqrt{3}\,  m \, \mu \,  \tilde{F}^{(0)}  
\Big]
\label{low} \eea
where  $\tilde{F}^{(0)} \equiv   { 1\o 2 }  \epsilon_{ij} F_{ij  }^0$  is the $U(1)$  flux. 
 This way one finds the nonabelian Bogomolny equations  ($\varepsilon=\pm1$),
\bea
&&  \frac1{2g_2 } F^{(a)}_{ij}+
     \frac{g_2}{4} \, \varepsilon \, 
\left(\bar{q}_A  S ^a q^A\right)   \epsilon_{ij}=0, \qquad  a=1,2,3;
\non \\
&&   \frac1{2g_1}F^{(0)}_{ij} +  
     \frac{g_1}{4\sqrt{3}} \, \varepsilon \, 
\left(|q^A|^2-2\xi \right)\epsilon_{ij}=0;
\non \\
 &&    \nabla_i \,q^A +i \, \varepsilon \, \epsilon_{ij} \, 
\nabla_j\, q^A=0, \qquad    A=1,2,\ldots, N_f .
\label{F38}  \eea
The properties of the  BPS   vortex solutions have been discussed in detail recently \cite{ABEKY}.
In fact,  there  exists a continuously degenerate family of vortex solutions of Eq.(\ref{F38}),     parametrized by $SU(2)_{C+F} /U(1)=CP^1 =S^2$.  
This is      due to the
system's exact  symmetry  $SU(2)_{C+F}\subset   SU(3)_c\times  SU(n_f)_F$ (remember $n_f=4,5$)
which is broken only by individual vortex configurations.
In \cite{ABEKY} it was also verified that such an exact symmetry is not spontaneously broken. In other words, the dual of the original $SU(2) \times U(1)$
theory in Higgs phase is indeed a confining     ${\cal N}=1 $  $SU(2)$  theory,   with two vacua!

{\it  This  implies the existence of  the corresponding  degenerate family of {\it monopoles}     which appear as sources of these vortices. 
For consistency,   the monopole  and vortex fluxes  must match precisely,  a fact to be proven      in Section  
\ref{sec:matching}  below.   }

\section {Monopoles and Vortices  Are Incompatible  \label{incomp}}

It might be tempting at this point to try to  search for a static solution of the nonabelian Bogomolny 
equations containing both the  vortex and the monopole.  However no such solution exists. 
 Monopoles are topologically stable  only if  $ \mu =
0:$   they represent
$\pi_2(SU(3)/(SU(2)\times U(1)/{\mathbb Z}_2))\sim \pi_1(SU(2)\times U(1)/{\mathbb Z}_2). $
At the low energy scales in which our $(SU(2)\times U(1))/{\mathbb Z}_2$ symmetry is entirely broken by the bare adjoint chiral multiplet mass, topologically nontrivial monopole configurations are classified by the homotopy group $\pi_2(SU(3))=0$ and thus the topological stability of our monopoles fails.

On the other hand vortices exist in the Higgs phase of the $H$ theory (which requires  $\mu \ne  0 $):  they represent the fundamental group
$\pi_1(SU(2)\times U(1)/{\mathbb Z}_2), $\footnote {
Note that the global symmetries, though important for explaining the appearance of the zero modes of these solitons or  of their flavor quantum numbers,
 do
not play  a role in  their stability. }        but the vortices are stable only approximately  
in the theory defined  at  scales much lower than  $ v_1$,  where monopole production is suppressed by a tiny  barrier penetration factor \cite{SY}.  
 
Mathematically,   the nonexistence of the vortex solution in the high-energy $SU(3)$ theory reflects  the fact that it  is  simply connected 
 ($\pi_1(SU(3))={\bf 0} $).    Physically, any vortex can  be attached to a monopole and antimonopole at the two ends:    clearly the
monopole-vortex-antimonopole configuration  cannot be a configuration of minimum energy,  as the energy decreases as  the vortex becomes shorter.

Summarizing,  the monopoles and vortices  are  incompatible  {\it as static configurations}.    This does not mean 
that it is incorrect to consider a vortex which ends on a monopole. Quite the contrary!   In fact, a (mesonlike) monopole-vortex-antimonopole  configurations
can rotate and can be dynamically  stable,  though as a static configuration they are not:  they do  not  representing  any nontrivial 
homotopy group element. After all, we believe that real-world mesons {\it are }  quark-gluon-antiquark   bound states  of this sort!

It is thus perfectly sensible to consider the physics   of ``a vortex ending on a monopole".  This notion will be made more quantitative and precise in the next section.

%but 
%we are obliged to  study  the sense of such a phrase  more precisely and quantitatively.  This is what is done in the next section.   

\section{Flux Matching  \label{sec:matching}  }

\begin{figure}[ht]  
\begin{center}
\leavevmode
\epsfxsize 12   cm
\epsffile{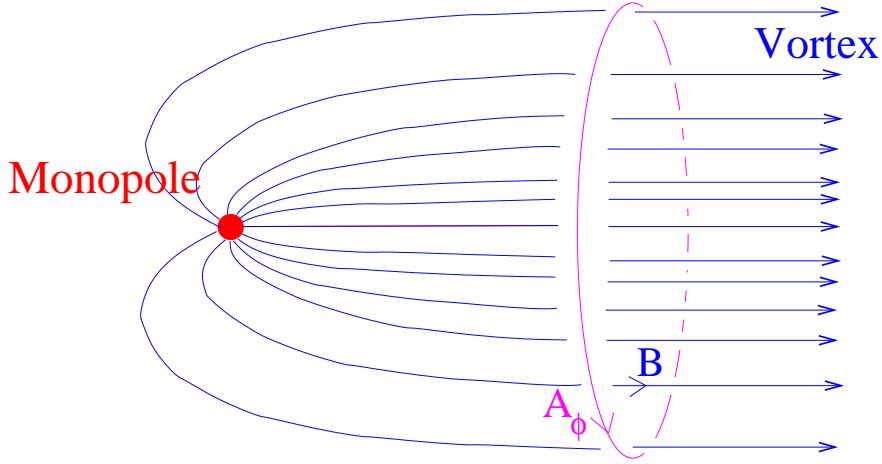}              
\end{center} 
\caption{A single vortex ends on each monopole.}
\label{monovort} 
\end{figure}

\begin{figure}[ht]  
\begin{center}
\leavevmode
\epsfxsize 12   cm
\epsffile{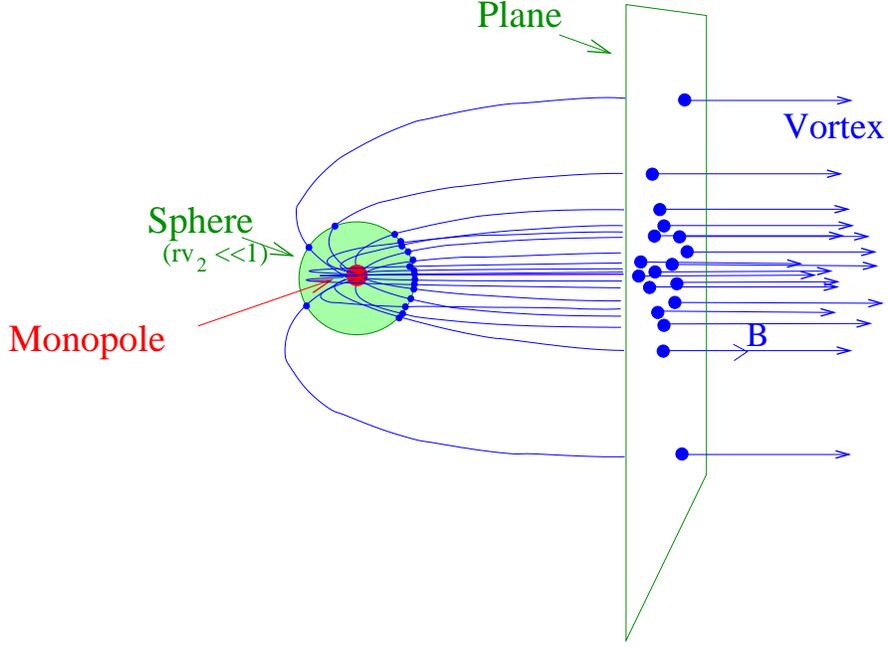}       
\end{center} 
\caption{{\small  The total flux around the monopole, integrated over a 
sphere of an arbitrary radius (hence with a radius much smaller than $1/v_2$ - where it lookes like an isotropic monopole),     must match the total vortex
flux
 integrated over a plane far enough from the monopole.  On this plane the vortex flux is
distributed over a region much larger than $1/v_2$.}}  
\label{fluxmatch}    
\end{figure}

Consider    the configuration in which a vortex ends on a monopole (Figure \ref{monovort}).  
The vortex and monopole both represent the same minimum element of 
\beq  \pi_2(SU(3)/(SU(2)\times U(1)/{\mathbb Z}_2))\sim\pi_1( {SU(2) \times U(1) \o {\mathbb Z}_2})= {\mathbb Z}.
\eeq
Therefore the  total flux through an ${\mathbb R}^2$  cross-section of the vortex and the total flux through an $S^2$  around 
the monopole on which it ends,  must agree  (Figure \ref{fluxmatch}).  In the rest of this section we shall  verify that this is indeed the case.  This means
that  when the $H$  theory is in Higgs phase the monopoles of the $G/H$  system are indeed confined.

As a bonus, we find  that the $U(1)$ charge of the nonabelian monopoles  takes a fractional value with respect to the
standard Dirac quantization condition, which is nicely  explained by  the homotopy group consideration   in Ref.~\cite{ABEKM}.

\subsection {Monopole flux}
Given the adjoint VEV
\beq \bra \phi \ket =
   \left( \begin{array}{ccc}
     v&0&0\\
     0&v&0\\
     0&0&-2 \, v\\
   \end{array} \right), 
\eeq
consider a  broken $SU(2)$ subgroup   (``$U$"-spin) with generators{\footnotesize\beq
  S_1\equiv  t^4= { 1\o 2}   \pmatrix{  0 & 0& 1  \cr  0 & 0 & 0 \cr  1  &0& 0   }; \quad  
 S_2\equiv t^5= { 1\o  2}   \pmatrix{  0 & 0& -i  \cr  0 & 0  & 0 \cr i &0& 0    }; \quad   S_3\equiv 
{ t^3 +  \sqrt3 t^8 \o 2} = { 1\o 2}   \pmatrix{  1  & 0  & 0  \cr  0 & 0 & 0 \cr  0&0& -1   }
\eeq}where $t^k$'s are  the Gell-Mann matrices.
A nonabelian monopole transforming in the doublet of the dual of this $SU(2)$ is described by the solution \cite{EW,BK}
\bea   \phi ({\bf r})  &=&
   \left( \begin{array}{ccc}
     -\frac{1}{2}v&0&0\\
     0&v&0\\
     0&0&-\frac{1}{2}v\\
   \end{array} \right)
   +
3\,  v\, {\vec  S} 
   \cdot \hat{r} \phi(r),   \non \\
  \vec{A}({\bf r}) &=& {\vec S} 
   \wedge \hat{r} A(r)      \label{su3sol}\eea
of the nonabelian Bogomolny equations (\ref{nbe}),
where   $\phi(r) \to 1$ and  $A(r) \to -{1 \o r}$.
Another, degenerate solution can be found by making use of the ``V-spin", lying in the $(2-3)$ submatrix.

The nonabelian flux is easily found to be
\beq   B_i =  { 1\o 2}  \epsilon_{ijk}  ( \de_j  A_k -  \de_k  A_j  - i   [A_j, A_k] )   %     \non \\
%&=&   { 1\o 2}  \epsilon_{ijk}   \Big( 2  \epsilon_{k \ell m} \de_j   {S_{\ell} r_m    \o r^2}  -   i  \epsilon_{j ts } \epsilon_{k \ell m} [S^t,  S^{\ell} ] { r_s r_m \o r^4}       
%\Big)  \non \\
%&=&( \delta_{i \ell} \delta_{jm} -   \delta_{i m} \delta_{j \ell } ) \Big( S_{\ell}  \, [    {\delta_{j m} -  2 \,  r_{j} r_m /r^2 \o r^2} ] +    { 1\o 2}  \epsilon_{j ts }
%\epsilon_{t\ell u} S^u        { r_s r_m \o r^4}    \Big) \non\\
%&=&( \delta_{i \ell} \delta_{jm} -   \delta_{i m} \delta_{j \ell } ) \Big( S_{\ell}  \,  [    {\delta_{j m} -  2 \,  r_{j} r_m /r^2 \o r^2}  ] -  { 1\o 2}   ( \delta_{j \ell}
%\delta_{su  } -  
%\delta_{j u} \delta_{ \ell s } )  S^u        { r_s r_m
%\o r^4}   
%\Big) \non \\
%&=&    2 \, {   r_i  ({\bf S}\cdot {\bf r} ) \o r^4}  -   {   r_i  ({\bf S}\cdot {\bf r}) \o r^4}  
=  {   r_i  ({\bf S}\cdot {\bf r})  \o r^4} . \label{calcu}
\eeq
To find the abelian flux along the unbroken $U(1)$, we project the nonabelian flux onto the direction of  the adjoint scalar and compute:
\beq  \Tr  \,  \phi  {\vec B}  = { 3 \o 2}   \, v  {\vec{r} \o r^3}. \label{so3flux}
\eeq
Integrating this over a 2-sphere centered on the magnetic monopole
and normalizing  the flux such that it is independent of the absolute value of the scalar condensate, $v$,   we find
\beq    F_m     =  \int_{S^2}  d{\bf S}  \cdot { \Tr  \,  \phi \,  {\bf  B} \o { 1\o \sqrt{2}}  (\Tr  \,  \phi \phi)^{1/2}   }  = 2 \pi   \cdot \sqrt{3  }.
\label{Monofluxbis} \eeq
We have chosen the normalization factor of 
\beq  { 1\o \sqrt{2}} ( \Tr  \,  \phi \phi)^{1/2}  =  { 1\o 2}  v  
\eeq
since
\beq    \Tr  \,  \phi \,  {\bf  B}  =  { 1\o 2} \, v \,  B^8.
\eeq

\subsection {Vortex flux}
Using the nonabelian vortex solution of \cite{ABEKY} 
\beq  \vec{A}_i  = t^8 \,  A^{8}_{i}(x)  + t^3 \, A^{3}_i(x) 
\eeq
\beq   A^{8}_{i}(x) = -\sqrt{3}\ \epsilon_{ij}\,\frac{x_j}{r^2}\,
[1-f_8(r)] \to -\sqrt{3} \,\epsilon_{ij}\,\frac{x_j}{r^2} = -\sqrt{3} \, { 1\o r}  \de_i  \varphi
\eeq
\beq
\label{phi}
\phi = -{1\over\sqrt{2}}\left(
\begin{array}{ccc}
  m & 0 & 0 \\
  0 & m & 0 \\
  0 & 0 & -2m
\end{array}\right) \equiv  \left(
\begin{array}{ccc}
  v & 0 & 0 \\
  0 & v & 0 \\
  0 & 0 & -2v
\end{array}\right) =  2 \sqrt{3}   \, t^8  \, v 
\eeq
we find that the flux carried by a vortex is
\beq   {\vec B} =  \nabla     \wedge   \vec{A}, \qquad   
%\eeq
 %\beq   F_v^{\prime}   =   \int d{\bf S}  \cdot \Tr  \,  \phi \,  {\bf  B}  = \int d{\bf S}  \cdot  \nabla     \wedge   \Tr  \,  \phi \,    \vec{A}    =  \int
%d {\bf {\ell}} \cdot   \Tr  \,  \phi \,    \vec{A}    =    
%-\sqrt{3} \cdot  2 \sqrt{3}    \cdot { 1\o 2}  \cdot 2\pi =   { 2 \pi    }     \cdot  3v \eeq
 F_v =   \int_{R^2}  d{\bf S}  \cdot { \Tr  \,  \phi \,  {\bf  B} \o { 1\o \sqrt{2}}  (\Tr  \,  \phi \phi)^{1/2}   }  = 
2 \pi   \cdot \sqrt{3  }. \label{Vorfluxbis} \eeq
The monopole flux (\ref{Monofluxbis}) and the vortex flux (\ref{Vorfluxbis})  agree precisely, and so  in our theory, in contrast with that of Ref.~\cite{Tong},
precisely one vortex ends on each monopole. 

\subsection{ Dirac} 

The magnetic flux sourced by a 't Hooft-Polyakov monopole \cite{TP,PS}    in a $SU(2) \to U(1)$ gauge theory is
 \beq    F_m     =  \int_{S^2}  d{\bf S}  \cdot {\bf  B}     = { 4 \pi \o   g  }, \qquad   g_{m}=   { 1 \o g}. 
\label{MonofluxTHP} \eeq
Here $g$ is the electric coupling constant, which enters the Lagrangian as
\beq  (\de_{\mu } -  i \, g {\tau^a \o 2   }A^a_{\mu}  )  \, q  
\eeq
where $q$ is an $SU(2)$ doublet matter field.  This means that the minimum electric charge is  $e_{0} =  { g \o 2}$ and so 
\beq    g_m=  { 1\o 2 \, e_0}
\eeq
coincides  with  Dirac's  minimum quantum of magnetic charge.

In the  $SU(3) \to SU(2) \times U(1)$  theory under consideration,   (\ref{Monofluxbis}) means that the minimum magnetic charge is 
\beq  g_{m}=   {\sqrt{3}  \o  2   g}. 
\eeq
But since  $g$ enters the Lagrangian as  
\beq     (\de_{\mu } -  i \, g {t^a }A^a_{\mu}  )  \, \pmatrix{q_1 \cr q_2 \cr q_3}  =     (\de_{\mu } -  i \, g {t^8 }A^8_{\mu} +
\ldots )  \, \pmatrix{q_1 \cr q_2 \cr q_3}  
\eeq
the minimum $A^8$ charge is 
\beq   e_0  =  { g \o 2 \sqrt{3}} \eeq
where the factor of $2\sqrt{3}$ comes from the normalization of the Gell-Mann matrix $t^8$.
 In terms of this,  the magnetic charge of the doublet monopole is 
\beq   g_m =  { 1 \o 4 \, e_0}
\eeq 
which is one half of the Dirac quantum. 
%{ \bf  As a result, in the presence of fundamental matter 
%our monopoles must be confined, although this calculation alone does not explain the
% fact that monopoles of charge 2 are also confined. }   %In string
% theory the confinement in the odd case comes from the orientability of the M2, but 
% then this might fail for SO and SP and so really the charge 2's could
% be free???

\subsection{Monopole charge and  flux in an  $SU(N+1)$  theory } 
The above analysis  generalizes straightforwardly to an $SU(N+1)$ gauge theory
 broken to $SU(N)\times U(1)$  by the adjoint scalar VEV
\beq    \phi =
 \pmatrix{  v     &  0  &  \ldots &    0 
  \cr  0   & v    & \ldots   & 0     
 \cr  \vdots  &\vdots& \ddots & \vdots       
\cr  0    & 0   &  \ldots  & -N v     } =
\pmatrix{  v \cdot {\bf 1}_{N\times N}     & 
  \cr    & -N v     } .
\label{phivev} \eeq
A magnetic monopole is characterized by the vector potential
\beq
\vec{A}(r)=\vec{S} \wedge \widehat{r}  \, \frac{A(r)}{g}
\eeq
which yields the magnetic flux 
\beq   B_i=   {   r_i  ({\bf S}\cdot {\bf r})  \o r^4}    
\eeq
where the matrices $S_i$  
{\small  \beq
S_1=\frac{1}{2} \pmatrix{  0     &  0  &  \ldots &    1 
  \cr  0   & 0    & \ldots   & 0  
 \cr  \vdots  &\vdots& \ddots & \vdots
\cr  1    & 0   &  \ldots  & 0     }; \qquad  
 S_2=\frac{1}{2} \pmatrix{  0     &  0  &  \ldots &    i
  \cr  0   & 0    & \ldots   & 0  
 \cr  \vdots  &\vdots& \ddots & \vdots
\cr  -i    & 0   &  \ldots  & 0     }; \eeq
\beq S_3=\frac{1}{2} \pmatrix{  1     &  0  &  \ldots &
   0 
  \cr  0   & 0    & \ldots   & 0  
 \cr  \vdots  &\vdots& \ddots & \vdots
\cr  0    & 0   &  \ldots  & -1     }
\eeq}
generate a broken $SU(2)$ subgroup of  $SU(N+1)$.  
In addition the adjoint Higgs field varies near the monopole as 
\beq
\phi=\pmatrix{  -\frac{N-1}{2} v     &  0  &  \ldots &    0
 &0
  \cr  0   & v    & 0 & \ldots    &0
\cr 0 & 0 & v & \ldots & 0
 \cr  \vdots  &\vdots& \vdots& \ddots & \vdots
\cr  0    & 0  &0 &  \ldots  & -\frac{N-1}{2} v   }
+(N+1) \, v \,  (\vec{S} \cdot \widehat{r}) \, \phi(r).
\eeq
Tracing in the $\phi$ direction, normalizing by the norm of $\phi$, and integrating over a sphere centered on the monopole,  we find the total magnetic flux
sourced by the minimal monopole   
\beq   F_m=  \int_{S^2}   d{\bf S}    \cdot {\Tr  \,  \phi \,  {\bf  B}  \o { 1\o \sqrt {2} }  ( \Tr \phi^2)^{1/2} }    %= { 2 \pi (N+1)  \o   \sqrt
%{N(N+1)} /\sqrt{2} } 
=  2 \pi \,
\sqrt{2(N+1)  
\o N}. 
\label  {Mflux}\eeq
This should be equal to $4 \pi \, g_m$, and so  
\beq   g_m=   \sqrt{N+1 \o  2 \, N} /g .
\eeq
On the other hand, the electric coupling of the $A^0_{\mu}$ field with the matter in the fundamental representation of $SU(N+1)$
is  through the minimum coupling constant
\beq  e_0=   { g  \o  \sqrt{2N(N+1)}},   
\eeq
as
\beq   t^0  =  { 1 \o  \sqrt{2N(N+1)}} 
\pmatrix{  {\bf 1}_{N\times N}     & 
  \cr    & -N      }. 
\eeq
Thus the minimum magnetic charge, in terms of the unit electric charge, is
\beq  g_m=   { 1 \o 2\,N\,  e_0}
\label{suncharge}\eeq
which is $1/N$  of the charge of Dirac's $U(1)$ monopole.
  This factor of $N$ is the degree of the embedding of the 
fundamental group of the unbroken $U(1)$ into that of the unbroken gauge group \cite{ABEKM}.

\subsection{Vortex flux in  the $SU(N+1)$  theory: Flux matching}

 The $(1,0,\ldots)$ - vortex solution  of  \cite{ABEKY} consists
 of  squark fields winding as 
\beq
 q^{kA}=\left(
 \begin{array}{ccc}
   e^{i\alpha}\phi_1 & 0 &0 \\
   0 & \ddots & 0 \\
   0 & 0 &   \phi_N \\  
   \end{array}\right)\label{vorq}
\eeq
while the adjoint scalar field  is fixed to its constant VEV. 
The vector potential cannot be found analytically, but instead is given in terms of the 
profile functions $f_i$ which solve a particular set of differential equations
$$
 A^3_{i}(x) = -\epsilon_{ij}\,\frac{x_j}{r^2}
 \Big(1 -f_3\Big),
$$
$$\vdots$$
$$
 A^{N^2 -1}_{i}(x) = -\sqrt{\frac2{N (N-1)}} \epsilon_{ij}\,\frac{x_j}{r^2}
 \Big((1 -f_{N^2-1}\Big),
$$
\beq    A_{i}(x) = -\frac1{  {\tilde e}   N} \, \epsilon_{ij}\,\frac{x_j}{r^2}
 \Big(1  -f\Big) =  - \frac1{ e }  \sqrt{ 2 (N + 1) \o N}  \, \epsilon_{ij}\,\frac{x_j}{r^2}  \label{vora}
\eeq¡
where we have rescaled 
\beq {\tilde e} \equiv  { e \o  \sqrt { 2 N ( 1 + N )}  };    \qquad  {\tilde A_i  } \equiv   { e   \o  {\tilde e} }   A_i.  \eeq
This may also be obtained directly directly  by solving    $(\nabla_i - A_i ) q \to 0$,  without the redefinition used in \cite{ABEKY}. 
Now  $$\phi =   \sqrt{2N(N+1)} \,  t^0  \, v$$
and so the abelian flux integrated over a cross-section of the vortex is 
\beq   F_v =  \int_{R^2}   d{\bf S}  \cdot    {   \Tr  \,  \phi \,  {\bf  B} \o { 1\o \sqrt {2} }  ( \Tr \phi^2)^{1/2} }    
  %{ 1 \o { 1\o \sqrt {2} }  ( \Tr \phi^2)^{1/2} }    \int_{C}     d{\bf \ell }  \cdot      \Tr  \,  \phi \,    \vec{A}    \non \\
%&=& \sqrt {  2 \o N  (N+1) } \, { 1\o 2} \,  \sqrt{2N(N+1)}  \, \sqrt{ 2 (N + 1) \o N} \, 2 \pi 
= { 2 \pi  }   \sqrt{2(N+1) \o N } , 
\eeq
in precise   agreement with the monopole flux of Eq.~(\ref{Mflux}).   We see again that one vortex ends on each monopole.

\subsection{Matching of   nonabelian fluxes}

We have seen above    that the abelian parts of the monopole and vortex fluxes agree.  In fact, the
full nonabelian fluxes must also match precisely, if computed in  the same  gauges. For the monopole, the solution is asymptotically
\beq
\phi=\pmatrix{  -\frac{n-1}{2} v     &  0  &  \ldots &    0
 &0
  \cr  0   & v    & 0 & \ldots    &0
\cr 0 & 0 & v & \ldots & 0
 \cr  \vdots  &\vdots& \vdots& \ddots & \vdots
\cr  0    & 0  &0 &  \ldots  & -\frac{n-1}{2} v   }
+(n+1) \, v \,  (\vec{S} \cdot \widehat{r})
\eeq
\beq
\vec{B}(r)= \frac{r_i({\bf S}\cdot {\bf r})}{r^4}.
\eeq
In  the  gauge in which $\phi$
is asymptotically 
\beq
 \pmatrix{  v     &  0  &  \ldots &    0
  \cr  0   & v    & \ldots   & 0
 \cr  \vdots  &\vdots& \ddots & \vdots
\cr  0    & 0   &  \ldots  & -n v     }
\eeq
in all spatial directions,    we find 
 \beq
\vec{B}(r)=\vec{S_3} \frac{r_i}{r^3}
\eeq
and therefore  the flux is given by:
\beq
\mathcal{F}_m=\int_{S^2}  d{\bf S}  \cdot {\bf B}= 4 \pi S_3=2 \pi \pmatrix{  1     &  0  &  \ldots &
   0
  \cr  0   & 0    & \ldots   & 0
 \cr  \vdots  &\vdots& \ddots & \vdots
\cr  0    & 0   &  \ldots  & -1     }. \label{mflux}
\eeq

 For the vortex, the solution given in  (\ref{vorq}),(\ref{vora}) is already in this    gauge:   $\phi$ is diagonal and constant.
The vortex flux is easily  found  to be
\beq
\mathcal{F}_v  =\int_{R^2}   d{\bf S}  \cdot {\bf B}=
\int_{C}   d{\bf l}  \cdot {\bf A} =2 \pi \pmatrix{  1     &  0  &  \ldots &
   0
  \cr  0   & 0    & \ldots   & 0
 \cr  \vdots  &\vdots& \ddots & \vdots
\cr  0    & 0   &  \ldots  & -1     },
\eeq
which precisely agrees  with    the monopole flux.

\section{Conclusion   \label{sec:quantum}}

 We have thus verified  that   in the theory with symmetry breaking 
\beq    SU(N+1)    \,\,\,{\stackrel {v_1}{\longrightarrow}}     \,\,\, {SU(N) \times U(1) \o  {\mathbb Z}_N  }    \,\,\,{\stackrel {v_2}{\longrightarrow}}     \,\,0,
\qquad  v_1 \gg v_2,
\eeq
the  massive monopoles  representing $\pi_2( { SU(N+1) \o SU(N) \times U(1) / {\mathbb Z}_N})$   are confined   by the nonabelian vortices
of the low-energy theory, which represent classes in  $\pi_1( {  SU(N) \times U(1) / {\mathbb Z}_N}).$  We have done so  by showing that the magnetic flux 
of one matches exactly that of the other.  As the two homotopy groups involved  are isomorphic, such an agreement might appear to be automatic, 
but the result is by no means trivial  since the monopoles and vortices are solutions  of different effective theories valid at different energy scales with 
different effective degrees of freedom. 
     
As a by-product, we have checked that the $U(1)$ charge of the monopoles is indeed ${ 1\o N}$ of the minimum Dirac quantum (for the  $U(1)$
 theory), a fact easily understood  from the minimum closed path in the space of ${  SU(N) \times U(1) / {\mathbb Z}_N} $ \cite{ABEKM}.

In the $SU(N+1)$    theories   discussed in this paper     there  are no other vortices or monopoles as both the first and second homotopy groups  are 
trivial.  When the original gauge group $G$ is not simply connected,  such as in the  ${SU(N) \o {\mathbb Z}_N}$  or  $SO(N)$ theories,  there are vortices in
the theory which are  sourced only by external (Dirac) monopoles.    These cases will be discussed elsewhere.

The most significant consequence of the flux matching discussion in this paper is the fact   that the  Goddard-Nuyts-Olive-Weinberg \cite{GNO,EW} 
monopoles  of the theory  indeed transform as the fundamental multiplet of the dual of $H$, ${\tilde H}=  SU(N) \times U(1) $.  This follows from the fact that 
 the  vortices of the  $H$ theory in the Higgs phase are described by a continuous family of degenerate  solutions (exact zero modes), parametrized by the 
quotient \cite{ABEKY}
\beq      {\mathbb {CP}}^{N-1}  \sim { SU(N)_{C+F}   \o  (SU(N-1)\times U(1) )_{C+F} }, 
\eeq
$(SU(N-1)\times U(1) )_{C+F} $   being the invariance group of an individual vortex. 
As the monopoles are the  sources of these vortices (nonlocal objects), it follows  that the monopoles themselves transform according to the continuous, 
dual transformations of 
${\tilde H}=  SU(N) \times U(1) $ which involve nonlocal field transformations.   The dual group $SU(N)$  also 
involves the original flavor subgroup 
$SU(N)\subset SU(N_f)$  in an essential manner. 

The flavor  symmetry group of the fundamental theory thus  plays {\it two }  crucial roles in the whole  discussion.  One is that through the effects of
the renormalization group the fermions  prevent  the unbroken   group $H$  from  becoming  strongly coupled and breaking    itself dynamically to an abelian
subgroup \footnote{This is precisely what happens in pure ${\cal N}=2 $ Yang-Mills theories or in a generic point of the moduli space of vacua.}.
Only in the  presence of an appropriate  number of massless  flavors \footnote{In our case,  the condition
is
$      2\, N +2  >  n_f 
\geq   2\, N.
$}   the  $H$  theory   remains  infrared free (or conformal invariant).   Otherwise,  the  ``nonabelian monopoles"  of the  bosonic theory   would remain 
simply    artifacts of the semi-classical  approximation \cite{BK}.   

Secondly,  the dual group ${\tilde H}$  itself   involves the original flavor group.     Such a mixing of  the groups of color and flavor 
is by now well-known  (if not so well understood) as exemplified  in the Seiberg   duals    occurring in many  ${\cal N}=1$   supersymmetric gauge
theory models
\cite{Sei}, and is really not surprising.

These  lessons  learned  from the supersymmetric world, though perhaps  not yet widely appreciated in the general physics community, 
might well be useful  in understanding the phenomenon of confinement in  the standard  QCD.

\section* {Acknowledgement}

We thank  N. Dorey,  H. Hansson, K. Higashijima, H. Murayama and A. Ritz    for discussions.


\begin{thebibliography}{100}

\bibitem{Lb}  E. Lubkin, {\bf Ann. Phys. 23}  (1963) 233.

\bibitem{WY}   T. T. Wu and  C. N. Yang,  {\bf  Phys. Rev. D12 }  (1975)  3845. 

\bibitem {MO} C. Montonen and D. Olive, {\bf Phys. Lett. 72 B} (1977) 117. 

\bibitem{GNO}   P. Goddard, J. Nuyts and D. Olive,   {\bf Nucl. Phys.  B125}
(1977) 1. 

\bibitem{SC}   S. Coleman, ``The  Magnetic Monopole  Fifty  Years Later,"
Lectures given at Int. Sch. of Subnuclear Phys., Erice, Italy  (1981). 

\bibitem{Chan} Chan Hong-Mo and Tsou Sheung Tsun,
   {\bf Nucl. Phys. B189}  (1981) 364.
 

\bibitem{EW}   E. J. Weinberg, {\bf Nucl. Phys. B167} (1980) 500;  {\bf Nucl. Phys. B203} (1982) 445. 

\bibitem{LWY} K. Lee, E. J. Weinberg and P. Yi,  {\bf Phys. Rev. D 54 } (1996) 6351, hep-th/9605229. . 

\bibitem{CDyons} A. Abouelsaood, {\bf Nucl. Phys. B226} (1983) 309; P. Nelson and A. Manohar,  {\bf Phys. Rev. Lett. 50}
(1983) 943; A. Balachandran, G. Marmo, M. Mukunda, J. Nilsson, E. Sudarshan and F. Zaccaria,   {\bf Phys. Rev. Lett. 50}
(1983) 1553;  P. Nelson and S. Coleman,  {\bf Nucl. Phys. B227} (1984) 1. 

\bibitem {CJH}   C. J. Houghton, P. M. Sutcliffe,
{\bf J.Math.Phys.38}  (1997) 5576, hep-th/9708006


\bibitem{Sei}  N. Seiberg, {\bf  Nucl.Phys.B435}  (1995) 129, hep-th/9411149;
   D. Kutasov, A. Schwimmer and N. Seiberg, {\bf  Nucl.Phys.B459}  (1996) 455, hep-th/9510222. 

\bibitem{APS}
P. C. Argyres, M. R. Plesser and N. Seiberg, Nucl. Phys. {\bf B471}
(1996)
159, hep-th/9603042;
P.C. Argyres, M.R. Plesser, and A.D. Shapere, {\bf 
Nucl. Phys.   B483}   (1997) 172,   hep-th/9608129.

\bibitem{CKM}
G. Carlino, K. Konishi and H. Murayama,
   {\bf  JHEP   0002}  (2000) 004,     hep-th/0001036;
 {\bf    Nucl. Phys.  B590}  (2000) 37,     hep-th/0005076.

\bibitem{CKKM}
G. Carlino, K. Konishi, Prem Kumar  and H. Murayama,
  {\bf    Nucl. Phys.  B608    }  (2001) 51, hep-th/0104064.


\bibitem{BK}   S. Bolognesi and K. Konishi,  {\bf    Nucl. Phys.  B645 }  (2002) 337, hep-th/0207161. 


\bibitem{ABEKY}   R. Auzzi, S. Bolognesi,  J. Evslin,  K. Konishi  and  A. Yung, to appear in {\bf Nucl. Phys. B}, hep-th/0307287.



\bibitem{HT}   A. Hanany, D. Tong,
hep-th/0306150.

\bibitem{Tong}   D. Tong,
hep-th/0307302.


\bibitem{SW1}
N. Seiberg and E. Witten,   {\bf   Nucl. Phys. B426} (1994) 19; Erratum
\textit{ibid.}     {\bf   Nucl.Phys.   B430} (1994) 485, hep-th/9407087.

\bibitem{SW2}
N. Seiberg and E. Witten, {\bf Nucl. Phys.  B431} (1994) 484,
   hep-th/9408099.

\bibitem{curves}

P.~C.~Argyres and A.~F.~Faraggi, {\bf Phys. Rev. Lett {\bf 74}} (1995)
3931, hep-th/9411047;
A. Klemm, W. Lerche, S. Theisen and S. Yankielowicz,  {\bf  Phys. Lett.
{\bf B344} }    (1995) 169, hep-th/9411048;
 {\bf  Int. J. Mod. Phys. A11}   (1996) 1929, hep-th/9505150;
A. Hanany
and Y. Oz,   {\bf  Nucl. Phys. {\bf B452} }   (1995) 283,
hep-th/9505075;
P.  C.  Argyres, M.  R.  Plesser and A.  D.  Shapere,   {\bf  Phys.  Rev.
Lett.  {\bf 75} }      (1995) 1699, hep-th/9505100;
P. C. Argyres and A. D. Shapere,   {\bf    Nucl. Phys. {\bf B461} }    (1996)
437,
hep-th/9509175;   A. Hanany,
 {\bf   Nucl.Phys. {\bf B466}}    (1996) 85,  hep-th/9509176.

\bibitem{SY}
M.~Shifman and A.~Yung,
{\bf Phys. Rev.  D66} (2002) 045012, hep-th/0205025. 

\bibitem{ABEKM}
R. Auzzi, S. Bolognesi,  J. Evslin, K. Konishi and H. Murayama,   in preparation.     


\bibitem{TP}  G. 't Hooft,  {\bf Nucl. Phys. B79} (1974) 276, 
  A. M. Polyakov,  {\bf JETP Lett. } 20 (1974)  194.

\bibitem{PS}  M. K. Prasad and C. M. Sommerfield, {\bf Phys. Rev. Lett. 35} (1075) 760;  S. Coleman, 
S. Parke, A. Neveu and C. M. Sommerfield, {\bf Phys. Rev.
D15} (1977)  544. 

\end{thebibliography}
\end{document}